\documentstyle[version2,preprint,aps]{revtex}

\begin{document}
\font\fortssbx=cmssbx10 scaled \magstep2
\hbox to \hsize{
\hskip.5in \raise.1in\hbox{\fortssbx University of Wisconsin -
Madison}
\hfill$\vcenter{\hbox{\bf MAD/PH/805}
                \hbox{Revised November 1994}}$ }
\vspace{.25in}
\begin{title}
{\bf Signatures of CP-violation in the Presence of Multiple \\
b-pair Production at Hadron Colliders}
\end{title}
\author{\large F.~Halzen$^1$, M.C.~Gonzalez-Garcia$^1$,
T.~Stelzer$^2$ and  R.A.~V\'azquez$^1$}
\begin{instit}
$^1$ Department of Physics, University of Wisconsin, Madison,  WI
53706, USA \\
$^2$ Department of Physics, University of Durham, Durham, UK.
\end{instit}
\thispagestyle{empty}

\begin{abstract}
We calculate the production of 2 b-quark pairs in hadron collisions. Sources
of multiple pairs are multiple interactions and higher order perturbative QCD
mechanisms. We subsequently investigate the competing effects of multiple
b-pair production on measurements of CP-violation: i) the increase in event
rate with multiple b-pair cross sections which may reach values of order 1
barn in the presence of multiple interactions and ii) the dilution of $b$
versus $\bar b$ tagging efficiency because of the presence of events with 4
$B$-mesons. The impact of multiple $B$-meson production is small unless the
cross section for producing a single pair exceeds 1~mb. We show that even for
larger values of the cross section the competing effects i) and ii) roughly
compensate so that there is no loss in the precision with which CP-violating
CKM angles can be determined.
\end{abstract}
\newpage
\section{Introduction}
CP-violation can be accommodated in the Standard Model with 3 families in terms
of a phase angle in the Cabibbo-Kobayashi-Maskawa (CKM) matrix. There are
several initiatives to study CP-violation in the $B$-sector of the CKM matrix.
Failure to observe the Standard Model predictions implies physics beyond the
Standard Model. It is customary to describe the $B$-sector of the CKM matrix
in terms of the  3 angles of the unitary triangle $\alpha$, $\beta$ and
$\gamma$  \cite{CPreview}. It is generally expected that a complete
determination of the 3 angles will require experiments at both $B$-factories
and hadron colliders.

In this paper we study the measurement of CP-violation angles at the LHC as a
function of its luminosity. The advantages and disadvantages of the collider
have been discussed in various publications \cite{Fridman}. The
LHC is projected to reach a peak luminosity of $10^{34}$ cm$^{-2}$ s$^{-1}$.
This combined with a large b-quark production cross section at high energy
guarantees a number of events of the order of $10^{10}$ per year. Even after
taking into account the branching ratios and detector efficiencies one is
still left with a high number of events, of the order of $10^4$. Therefore the
LHC is a $B$-factory.

The highest luminosities are, however, reached at a price: the possible
presence of multiple interactions per beam crossing which could lead to the
production of multiple b-quark pairs in the detector. The multiple b-pair
cross section increases as the number of interactions per beam
crossing and may reach 1 barn at the highest luminosity where it dominates the
single pair cross section.  In extreme cases this effect may increase the
number of b-pairs produced by two orders of magnitude.  Unfortunately, double
pair production will also introduce an additional source of fake asymmetry.
In a double pair event   the gold-plated tagging {\it e.g.} $B^0
B^- \rightarrow J/ \psi K_s l^- X$ as well as the abnormal pairing {\it e.g.}
$B^0 B^+  \rightarrow J/ \psi K_s l^+ X$ are possible. Therefore the number of
mistagged b's increases since there is no possibility of tagging
all four $b$-hadrons in the event.

We will in the end conclude that the impact of multiple $B$-meson production
is small unless the cross section for producing a single pair exceeds 1~mb.
Even in this case, the competing effects of i) a gain in event rates from the
additional production of multiple b-pair events  and ii) the dilution of the
CP-signature by the introduction of fake asymmetry associated with wrong
pairing of b's, are likely to cancel. If one efficiently controls the
systematics of the mistagging, the sensitivity of the experiments is actually
improved. We will present results quantifying these statements.

On the theoretical side, we compute the production of 4 b-quarks in a single
pp-interaction to leading order in QCD. Its cross section is comparable to the
cross section for producing 2 pairs of b-quarks by multiple partons
when no overlapping events occur.

\section{Single Versus Double Pair Production}

At high energies events with more than one heavy quark pair become abundant
\cite{Keeble,Halzen_3}. In a hadron collider event there are 3 sources of
events with 2 b-quark pairs: i) higher order QCD diagrams, ii) double parton
interaction in a single pp collision and, iii) multiple pp interactions in the
crossing of the two bunches. For high luminosity LHC running mechanisms ii) and
iii) dominate and the cross section for the production of 2 b-quark pairs can
be written as
\begin{equation}
\sigma_{4b}= \sigma_{2 b}^2 \left\{ \frac{1}{2\pi R^2} +
2\sqrt{\frac{1}{2\pi R^2}\frac{(N-1)}{\sigma_{inel}}} +
\frac{(N-1)}{\sigma_{inel}}  \right\}.
\label{MP}
\end{equation}
The 3 terms in Eq. \ref{MP} correspond, respectively, to the 3 diagrams in
Fig.\ref{diagrams}. The first term is the cross section for double parton
interaction as estimated in \cite{Goebel}.
$\sigma_{2 b}$ is the single b-pair production cross section. In  our
calculation we include $\sigma_{2b}$ from perturbative QCD to order
$O(\alpha_s^3)$ \cite{Ellis}. $R$ is the radius of the proton occupied by the
partons,
mostly gluons, producing the b-pairs (in our calculation we use $R=1$ fm).
$N$ is the average number of interactions per bunch crossing,
\begin{equation}
N= {\cal L} \Delta t \sigma_{inel}.
\end{equation}
where ${\cal L}$ is the luminosity and $\Delta t$ the time between collisions.

The last term in Eq. \ref{MP} is the standard form for the cross section for
multiple  p-p interaction in the bunch crossing with $\sigma_{inel}$ being the
proton-proton inelastic cross section. As for the second term in Eq. \ref{MP},
this is the cross section for the interaction of two partons from one proton in
one bunch with two protons from the other bunch. It can be understood as the
interference of the two previous effects.
{}From Eq. \ref{MP} we see that $\sigma_{4 b}$ increases linearly with the
luminosity. We present our results on cross sections next.

In Fig.\ref{cross}.a the single and double $b \bar b$ pair production cross
sections in $p~p$ collisions are shown as a function of $\sqrt{s}$.  As seen in
the figure, at high energy, production by multiple parton interaction
dominates.

For the sake comparison we also show the cross section for the the higher-order
QCD process $pp\rightarrow b\bar b b\bar b$. Its calculation is involved since
it receives contributions from 72 Feynman diagrams. Our results are in good
agreement with  the leading-logarithm calculation of reference \cite{Halzen_3}.
Fig. \ref{distributions}  shows several of characteristics of these events.
The invariant energy of the interaction is around 100 GeV which corresponds to
a Bjorken $x$ of about  0.07.  Since gluon distributions are well tested at
this energy we expect  our results to be fairly insensitive to the choice of
parton distributions. The rapidity, transverse energy, and $\eta$--$\phi$
separation show that the b pairs are experimentally accessible. However the
cross section is two orders of magnitude smaller than the single b-pair
production cross section.

As for the multiple parton interaction events, one should expect them to have
similar distributions as the ones from single events since they are not more
than an overlapping of those.

Figure \ref{cross}.b shows the double pair production cross section by multiple
parton interactions as a function of the LHC luminosity for 2 values of the
single pair cross section. The values bracket the expectations which strongly
depend on the gluon structure function. Multiple parton  processes may dominate
single pair production for ${\cal L}> 2-20\times 10^{33} $cm$^{-2}$ s$^{-1}$
for $\sigma_{2b}>1-10$ mb.

It should be pointed out that one cannot expect a to make a reliable estimate
of $\sigma_{2b}$ at the LHC from leading order perturbation theory. The leading
order perturbative calculations, which require inclusion of $O(\alpha_s^2)$ and
$O(\alpha_s^3)$ diagrams, are unreliable because the results are sensitive to
the assumed quark mass and the renormalization scale. They seem to already
underestimate the experimentally observed cross section at the Tevatron
\cite{teva}. Because of its intermediate mass the calculation of the bottom
quark cross section is believed not to be well understood. A perturbative
calculation lies beyond the scope of existing QCD technology because it
requires resummation of large logarithms of $1/x$ with $x\simeq
m_b/\sqrt{s}$\cite{collins}. The range of values, considered in this paper, is
conservative but cannot be guaranteed.

\section{CP Violation Measurement. Results }

We will illustrate the implications of multiple b-pair production for the
study of CP-violation using the gold plated channel $\bar{B}^0_d
\rightarrow J/ \psi K_s$ followed by  $K_s\rightarrow \pi^+\pi^-$ and
$J/ \psi\rightarrow l^+l^-$, which yields information on the angle $\beta$.
The $b$ or $\bar b$ nature of the second $B$ is established by the sign of
the lepton produced in its semileptonic decay.
CP-violation results in a non-vanishing asymmetry:
\begin{equation}
A=\frac{N_+-N_-}{N_++N_-},
\end{equation}
where $N_{+,-}$ represent the number of events of the type $B \rightarrow
J/\psi K_s , \bar{B} \rightarrow J /\psi K_s$, respectively. The integrated
asymmetry is given by
\begin{equation}
A=\frac{x_d}{1+x_d^2} \sin (2 \beta),
\end{equation}
where $x_d=\Delta m /\Gamma \; \sim 0.69 $, $\Delta m$ is the mass difference
between the two states and $\Gamma$ is their width. The factor
$\frac{x_d}{1+x_d^2}$ accounts for the possibility of $B^0_d$ oscillation.

For proton-proton interactions the initial state is not a CP eigenstate, so one
expects to have a fake contribution to the asymmetry, $F$, which modifies the
asymmetry to
\begin{equation}
A \simeq a[ \sin (2 \beta) + F],
\label{asym}
\end{equation}
with $a= D \frac{x_d}{1+x_d^2}$. $D$ is related to the so-called ``dilution''
factor. The equation is valid for small $\beta$. $F$ has been estimated to be
of order of  a few percent of the signal \cite{DeRujula}.

We are now ready to analyze the effect of multiple parton interactions on the
measurement of CP-violation at LHC. When including double pair production we
must account for all possible pairings. We will assume that the trigger is
designed to tag a single $B$-hadron. In the presence of multiple pair
production the ``other'' $B$-hadron may, or may not, be assigned the
``correct'' charge. This will increase the fake contribution to the asymmetry.

Let us define $\alpha$ as the ratio of double to single pair
production;
\begin{equation}
\alpha=p \frac{\sigma_{4b}}{\sigma_{2 b}}= p \left\{ \frac{\sigma_{2 b}}{2\pi
R^2}+ (N-1) \frac{\sigma_{2 b}}{\sigma_{inel}} +2\sigma_{2
b}\sqrt{\frac{1}{2\pi R^2}\frac{(N-1)}{\sigma_{inel}}} \right\},
\end{equation}
where $p \sim 0.64$ is the probability that two or more $B$'s do not decay
semileptonically since we are assuming only one prompt lepton in the event.
Then the asymmetry can be written as:
\begin{equation}
A=\frac{N^1_+ + \alpha N^2_+ - N^1_- - \alpha N^2_-}
       {N^1_+ + \alpha N^2_+ + N^1_- + \alpha N^2_-},
\label{asymmetry}
\end{equation}
where $N^{1,2}_{+,-}$ represent the number of events produced with $+$ or $-$
signature and coming from a double ($2$) or single ($1$) $b$-pair production.
We have that
\begin{equation}
\begin{array}{lllll}
N_+^{i=1,2}=N(J/\psi \; K_s \; l^+ X)^{1,2} & \propto
       & \{ \tilde{n}^+ & + R_d \tilde{n}^0 + R_s \tilde{n}^s \} &
         \{ 1 + \frac{x_d}{1+x_d^2} \sin 2 \beta \} \nonumber \\
  & + &  \{  & W_d \tilde{n}^0 + W_{\bar s} \tilde{n}^{\bar s} \} &
         \{1 - \frac{x_d}{1+x_d^2} \sin 2\beta \} \nonumber \\
  & + &  \delta_{i,2}\{ n^+ & + R_d n^0 + R_{s} n^{s} \} &
         \{1 - \frac{x_d}{1+x_d^2} \sin 2\beta \} \nonumber \\
  & + &  \delta_{i,2}\{  & W_d n^0 + W_{s} n^{s} \} &
         \{1 + \frac{x_d}{1+x_d^2} \sin 2\beta \} \nonumber \\
\end{array}
\end{equation}
\begin{equation}
\begin{array}{lllll}
N_-^{i=1,2}=N(J/\psi \; K_s \; l^- X) & \propto
       & \{ \tilde{n}^- & + R_d \tilde{n}^0 + R_{s} \tilde{n}^{\bar  s}\} &
         \{1 -\frac{x_d}{1+x_d^2} \sin 2\beta \} \nonumber \\
  & + &  \{  & W_d \tilde{n}^0 + W_{s} \tilde{n}^{\bar s} \} &
         \{1 + \frac{x_d}{1+x_d^2} \sin 2\beta \} \nonumber \\
  & + &  \delta_{i,2}\{ n^- & + R_d n^0 + R_{s} n^{\bar s} \} &
         \{1 + \frac{x_d}{1+x_d^2} \sin 2\beta \} \nonumber \\
  & + &  \delta_{i,2}\{  & W_d n^0 + W_{s} n^{\bar s} \} &
         \{1 - \frac{x_d}{1+x_d^2} \sin 2\beta \} \nonumber \\
\end{array}
\label{Npm}
\end{equation}
The effect of neutral $B$ oscillation in the tagging is parametrized in
terms of the $W_i$ and $R_i=1-W_i$ coefficients. $W_i$ is the probability
of a $B^0_i$ meson to decay as a $\bar B^0_i$
\begin{equation}
W_i=\frac{x_i^2}{2+2x_i^2}
\end{equation}
We have used $x_d=0.69$ and $x_s=9$ \cite{DeRujula}

The terms proportional to $\delta_{i,2}$ account for the possibility of
abnormal pairing in the double pair processes. We used the same notation as in
reference \cite{DeRujula}:
\begin{equation}
\begin{array}{lll}
\tilde{n}^+= & 2 C [ N(B^+ \bar{B}^0) Br(B^+ \rightarrow l^+) + &
                     N(\bar{\Lambda}_b \bar{B}^0) Br(\bar{\Lambda}_b
                   \rightarrow l^+) ] \\
\tilde{n}^-= & 2 C [ N(B^- \bar{B}^0) Br(B^- \rightarrow l^-) + &
                     N(\Lambda_b \bar{B}^0) Br(\Lambda_b
                   \rightarrow l^-) ] \\
\tilde{n}^0= & C N(B^0 \bar{B}^0) [Br(B^0 \rightarrow l^+) + &
                 Br(\bar{B}^0 \rightarrow l^-)] \\
\tilde{n}^s= & 2 C N(B^0_s \bar{B}^0) Br(B^0_s \rightarrow l^+ ) & \\
\tilde{n}^{\bar s}= & 2 C N(B^0_{\bar s} \bar{B}^0)
Br(B^0_{\bar s} \rightarrow l^- ) & \\
\end{array}
\end{equation}
\begin{equation}
\begin{array}{lll}
n^+= & C [ N(B^+ B^0) Br(B^+ \rightarrow  l^+)  + &
        N(\bar{\Lambda}_b B^0)  Br(\bar{\Lambda}_b \rightarrow l^+) ]
\\
n^-= & C [ N(B^- \bar{B}^0) Br(B^- \rightarrow  l^-)  + &
        N(\Lambda_b \bar{B}^0)  Br(\Lambda_b \rightarrow l^-) ] \\
n^0= & C [ N(B^0 B^0) Br(B^0 \rightarrow  l^+)  + &
        N(\bar{B}^0 \bar{B}^0)  Br(\bar{B}^0 \rightarrow l^-) ] \\
n^s= & C  N(B_s^0 B^0) Br(B_s^0 \rightarrow  l^+) & \\
n^{\bar s}= & C N(\bar{B}_s^0 \bar{B}^0)  Br(\bar{B}_s^0 \rightarrow
l^-) & \\
\end{array}
\end{equation}
Here $\tilde{n}^{\alpha}$ and $n^{\alpha}$, $\alpha=+,-,0,s,\bar{s}$
correspond to ``favorable'' and ``unfavorable'' assignments, respectively. $C$
is a constant. Even after including all possibilities the asymmetry can still
be cast in the form of Eq. \ref{asym} for small $\sin(2 \beta)$. $F$ and $a$
must, however, include the contributions from multiple pair production.

Because the initial state is not a CP eigenstate the  $N(B_i  \bar B_j)$ are
not all equal since protons are made of quarks rather than antiquarks and
therefore a $\bar b$ is more likely to pair with a valence quark than a $b$.
This difference is parametrized \cite{DeRujula}  in terms of  baryon and meson
fragmentation probabilities $l_b, l_m$ which are implicitly defined by the
equations:
\begin{equation}
N(B_i  \bar B_j) \propto P(B_i) P(\bar B_j)
\end{equation}
with
\begin{equation}
\begin{array} {llll}
P(B^+)=p_d(1-l_m)+2l_m/3 & & & P(B^-)=p_d(1-l_b) \\
P(B^0_d)=p_d(1-l_m)+l_m/3 & & & P(\bar B^0_d)=p_d(1-l_b) \\
P(B^0_s)=p_s(1-l_m) & & & P(\bar B^0_s)=p_s(1-l_b) \\
P(\bar \Lambda_b)=p_\Lambda(1-l_m) & & & P(\Lambda_b)=p_\Lambda
(1-l_b)+l_b\\
\end{array}
\end{equation}
We will assume that $p_d=0.38$, $ p_s=0.14$ and $p_\lambda=0.1$ and take the
semileptonic branching ratios for $\Lambda_b$ and $B$ to be equal
\cite{DeRujula}.
Clearly for $l_m=l_b=0$ it follows that $N(B_i \bar B_j)=N(B_j \bar B_i)$ and
the fake asymmetry vanishes. The results are not very sensitive to $ l_m$ but
depend, in contrast, very strongly on the value of $l_b$.

In Figs. \ref{A(0)} , \ref{Fake} and \ref{dilu} we have displayed the value of
$A(0)$,  the coefficient $a$, and the fake contribution to the asymmetry, $F$,
as a function of the LHC luminosity for different values of $\sigma_{2b}$,
$\sigma_{inel}$ and the parameters $l_b$ and $l_m$. As expected the values of
$A(0)$ and $F$ depend strongly on the parameter $l_b$ while $a$, which
measures the ``true'' final state asymmetry, is rather insensitive to it. As
for the effect of multiple pair production, it increases the value of the fake
contribution to the asymmetry while decreasing the value of the $a$. As
expected, it increases the ratio ratio of ``fake'' to ``true'' asymmetry. The
value of the luminosity at which this effect becomes important depends on the
value of the $\sigma_{2b}$. For $\sigma_{2b}< 1$ mb the effect is not very
important for any possible value of the luminosity.

To quantify the impact of double pair-production on the measurement of
$\sin(2\beta)$ we have computed the statistical and systematic error on
$\sin(2 \beta)$, with
\begin{equation}
\sigma_{stat}= \frac{\sqrt{1 -A^2}}{\sqrt{2 N_{events}}}
\end{equation}
\begin{equation}
\sigma_{syst}=\sqrt{\sin^2(2 \beta) \sigma^2_{a}/a^2 + \sigma^2_{F}},
\end{equation}
where $N_{events}$ is the total number of signal events given by
\begin{equation}
\begin{array}{ll}
N_{events} &= N(J/\psi K_s l^+)+N(J/\psi K_s l^-) \\
 & = 2 \times N(b\bar b)\times P(B_d) \times Br(B\rightarrow l X)
\times Br(B_d\rightarrow J/\psi K_s) \\ & \times Br (K_s\rightarrow
\pi^+\pi^-)\times Br(J/\psi \rightarrow l^+l^-)
\times \epsilon \\
 & = 6.76 \times 10^6 \times ({\cal L}/10^{34})\times
(\sigma_{2b}/\mbox{mb})\times (1+\alpha)\;\; \mbox{events/year}
\end{array}
\end{equation}

The $\epsilon$-factor is the product of the triggering, tagging and
reconstruction efficiencies combined with the geometrical acceptance. We use
$\epsilon=7.7\times 10^{-3}$ \cite{Erhan}. The systematic errors on $F$ and
$a$ have been estimated to be of the order of a percent \cite{Atlas}.

In summary, the presence of multiple pair production increases the total
number of events and therefore improves the statistical error. It also
increases the fake contribution to the asymmetry, $F$, and, as a result of
this, it increases the systematic error. The minimum value of $\sin(2\beta)$
that can be measured  with a given statistical significance $N_\sigma$ is
\begin{equation}
\sin(2\beta)_{min}=N_{\sigma} \frac{\displaystyle
\sqrt{\sigma^2_{stat}+\sigma_F^2}} {\displaystyle
\sqrt{1-N_\sigma^2\frac{\sigma_a^2}{a^2}}}
\label{sinmi}
\end{equation}

In Fig. \ref{sinmin}  we plot $\sin(2\beta)_{min}$ for $\sigma_{2b}=10$ mb as
a function of the collider luminosity for different values of the systematic
errors on $F$ and $a$ for a  3$\sigma$ effect. For comparison we show the
corresponding value of $\sin(2\beta)_{min}$ for $\sigma_{2b}=0.1$ mb. As
expected the value of $\beta_{min}$ and its dependence on luminosity depend
critically on the systematic error on the fake asymmetry and are less
sensitive to other parameters such as the systematic error on $a$. This is
easy to understand.  Unless the systematic error on $F$ is extremely small,
its systematic error represents the limiting factor on $\sin(2\beta)_{min}$ in
Eq. \ref{sinmi}. For large values of $\sigma_{2b}$, the presence of multiple
pair production increases $F$ at higher luminosities and therefore the value
of the $\sin(2\beta)_{min}$. If, on the other hand, one can determine $F$ with
good precision, the limiting factor in Eq. \ref{sinmi} becomes the statistical
error and the increased statistics as a result of multiple b-pair production
improves the measurement.

\section{Conclusions}
In this paper we have studied CP-violation measurements at hadron colliders.
We have investigated the effect of multiple b-production as a function of
collider luminosity. We conclude that  multiple b-pair production dominates
over single pair production for ${\cal L}> 2-20\times 10^{33} $cm$^{-2}$
s$^{-1}$ for $\sigma_{2b}>1-10$ mb.

The presence of multiple pair production produces two competing effects in the
measurement of the CP violating asymmetry. It increases the total number of
events and therefore it improves the statistics. It also increases the number
of mistagged events introducing an additional source of fake asymmetry and
therefore worsens the systematics.

In the end the impact of multiple $B$-meson production is small unless the
cross section for producing a single pair exceeds 1~mb. In this case the
dominant factor weighing the competing effects is the error on the
determination of the fake contribution to the asymmetry. Lower luminosities
are advantageous when one cannot measure the fake contribution to better than
a few percent. However, if a higher precision is achieved, it is still
advantageous to perform the measurement at a higher luminosity.

\acknowledgments
We thank A. Nisati for very helpful discussions.
This work was supported by the University of Wisconsin Research Committee with
funds granted by the Wisconsin Alumni Research Foundation, by the
U.S.~Department of Energy under Contract No.~DE-AC02-76ER00881 and by the
Texas National Research Laboratory Commission under Grant No.~RGFY9273.

\figure{Diagrams
for double pair production by multiple parton interaction in high energy $pp$
collisions. The three contributions correspond to the  three terms in Eq.
\ref{MP} {\bf a)} $\sigma_{2 b}^2 \frac{1}{2\pi R^2}$, {\bf b)} $\sigma_{2 b}^2
\sqrt{\frac{1}{2\pi R^2}\frac{(N-1)}{\sigma_{inel}}}$, and {\bf c)} $\sigma_{2
b}^2 \frac{(N-1)}{\sigma_{inel}}$. \label{diagrams}}
\figure{{\bf a)} Single
and double $b \bar b$ pair production cross sections in $p~p$ collisions as a
function of $\sqrt{s}$. The shaded band represents the prediction for the
single pair production cross section $\sigma_{2b}$ from perturbative QCD to
order $O(\alpha_s^3)$ obtained for several parametrizations of proton structure
functions and different values of the scale $\mu$ and the quark mass $m_b$. The
upper (lower) edge of the band corresponds to the prediction for the
parametrization in Ref. \cite{Nason} (EHLQ \cite{EHLQ}) for the structure
functions with $m_b=4.5\; (5.0)$ GeV and $\mu=mb/2\;(2 m_b)$. $\sigma_{4bPERT}$
(dashed lines) is the cross section from higher-order QCD processes
$pp\rightarrow b\bar b b\bar b$ for the EHLQ structure functions with $m_b=5$
GeV. The upper (lower) curve corresponds to $\mu=4 m_b \;(\sqrt{\hat s})$.
$\sigma_{4bMP}$ (full lines) is the double pair production cross section from
multiple parton processes (Eq.\ref{MP}). The lower (upper) curve corresponds to
$N=1\; (50)$ and the lower (upper) values of $\sigma_{2b}$. We have used the
prediction of $\sigma_{inel}$ from \cite{QGSM}. Here N is the number of
interactions per bunch crossing.\\ {\bf b)} Double b-pair production cross
section from multiple parton processes (Eq.\ref{MP}) at LHC ($\sqrt{s}=16$ TeV)
as a function of luminosity for $\sigma_{inel}(\sqrt{s}=16$ TeV)=85 mb
\cite{QGSM} and $\Delta t=25$ ns. The upper (lower) curve corresponds to
$\sigma_{2b}=10~(0.1)$ mb. \label{cross}}
\figure{Invariant mass ($\sqrt{\hat s}$), pseudorapidity ($\eta$), transverse
energy, and  $\eta$--$\phi$ separation  ($\Delta r$) distributions for the
4-b's events from higher order QCD diagrams. The figures are shown for EHLQ
structure functions with $\mu=4 m_b$. As discuss in the text the distributions
are not very sensitive to this choice. \label{distributions}}
\figure{Asymmetry for $\sin(2\beta)=0$,  $A(0)$, at LHC as a function of the
collider luminosity for two values of $\sigma_{2b}$. The upper curves
correspond to $l_b=0.$ and $l_m=0.05$. The central curves correspond to
$l_b=l_m=0.05$. and the lower curves correspond to $l_b=0.1$ and $l_m=0.05$. In
all cases the dark (light) lines correspond to $\sigma_{inel}=100~ (50)$ mb.
\label {A(0)}}
\figure{Same as previous figure for the fake contribution to
the asymmetry, $F$. \label{Fake}}
\figure{Same as previous figure for the
coefficient $a$. \label{dilu}}
\figure{Minimum value of $\sin(2\beta)$
measurable at the LHC with 3$\sigma$ effect as a function of the collider
luminosity for $\sigma_{2b}=10$ mb (solid lines ) and $\sigma_{2b}=0.1$ mb
(dotted lines), $\sigma_{inel}=100$ mb, and $l_m=l_b=0.05$. From upper to lower
the curves correspond to $\sigma_a=\sigma_F=30~,10~,~1~$. \\ \label{sinmin}}
\end{document}